\documentclass{ws-ijmpe}

\usepackage{graphics}
\usepackage{epsfig}

\begin{document}

%\markboth{V.Riabov}{PNPI (Measurement of the multi-hadron decays of $\omega$, $K_{S}^{0}$ and $\eta$-mesons in heavy ion collisions at $\sqrt{s_{NN}}=200$~$GeV$ in the PHENIX experiment at RHIC)}

\catchline{}{}{}{}{}
%%%%%%%%%%%%%%%%%%%%%%%%%%%%%%%%%%%%%%%%%%%%%%%%%%%%%%%%%%%%%%%%%%%%

\title{Measurement of the multi-hadron decays of $\omega$, $K_{S}^{0}$ and $\eta$-mesons in heavy ion collisions at $\sqrt{s_{NN}}=200$~$GeV$ in the PHENIX experiment at RHIC}

\author{\footnotesize V.Ryabov (for the PHENIX Collaboration)}

\address{Petersburg Nuclear Physics Institute, 188300, Russia, Gatchina \\
riabovvg@mail.pnpi.spb.ru}

\maketitle

\begin{history}
\received{(received date)}
\revised{(revised date)}
%\accepted{(Day Month Year)}
%\comby{(xxxxxxxxxx)}
\end{history}

\begin{abstract}
The PHENIX experiment at RHIC measured $\omega$, $\eta$ and $K_{S}^{0}$-meson production at high $p_{T}$ in $p+p$, $d+Au$ and $Au+Au$ collisions at $\sqrt{s_{NN}}=200$~$GeV$. Measurements performed in different hadronic decay channels give consistent results. The measured ratios of all three mesons to $\pi^{0}$ are found to be flat as a function of $p_{T}$ in $p+p$ collisions and equal to $\omega/\pi^{0}=0.81\pm0.02\pm0.07$, $\eta/\pi^{0}=0.48\pm0.02\pm0.02$ and $K_{S}^{0}/\pi^{0}=0.45\pm0.01\pm0.05$. Nuclear modification factor measured for $\omega$-mesons in central Au+Au collisions is $R_{AA}=0.4\pm0.15$. 
\end{abstract}

\section{Introduction}
Hadron spectra measured at high transverse momentum ($p_{T} > 2$~$GeV/c$) provide important information on particle production mechanism and properties of the matter produced in RHIC collisions. High-$p_{T}$ hadrons are mainly produced by fragmentation of partons originating from hard scattering processes. In nucleon-nucleon collisions such processes are relatively well understood and corresponding cross sections can be calculated in perturbative QCD. In nucleus-nucleus collisions various nuclear effects like multiple scattering in the initial state, modification of the parton distribution and fragmentation functions, energy loss in the medium can modify particle production. These effects can be studied by comparison of particle properties measured in $p+p$ and nucleus-nucleus collisions at the same energy. Measurement of $\omega$, $\eta$ and $K_{S}^{0}$-meson production in $p+p$, $d+Au$ and $Au+Au$ collisions is a part of systematic study of particle properties at RHIC.

\section{Experimental setup and data samples}
Two central spectrometers of the PHENIX~\cite{phenix} experiment ($\Delta\phi <$ $90^{0}$, $\mid\eta\mid < 0.35$) are used to measure neutral and charged particles produced in heavy ion collisions at RHIC. Beam-Beam Counters and Zero Degree Calorimeters provide the minimum bias (MB) trigger, determine $z$-coordinate of the collision vertex and the event centrality. Drift Chamber and the first layer of the Pad Chamber are used for reconstruction of charged particle momentum with resolution of $\sigma(p_{T})/p_{T} \approx 1.0$\%$p_{T}\oplus1.1\%$. Two other layers of the Pad Chamber can be used for track confirmation by matching it to the reconstructed hit. High segmentation electromagnetic calorimeter (EMCal) with energy resolution of $\sigma(E)/E \approx 8.1$\%$/\sqrt{E}\oplus2.1$\% is used as a primary detector for reconstruction of $\gamma$-quanta and $\pi^{0}$-mesons~\cite{neutmes}. 

The experimental data samples gathered by PHENIX in years 2003-2005 and decays analyzed in these samples are summarized in Table~\ref{tab:analdec}. Besides the  MB trigger we use high-$p_{T}$ $\gamma$-trigger ($\gamma$) realized by adding together amplitudes in $4\times4$ adjacent EMCal towers and comparing them to a threshold of 1.4~$GeV$ in $p + p$ and 2.4~$GeV$ in $d+Au$ collisions. Only events with the vertex $\mid$$z_{vert}$$\mid < 30$ cm from the center of the PHENIX experimental region and satisfying various quality assurance cuts are accepted in the analysis. 
\begin{table}[h]
\tbl{\label{tab:analdec}Analyzed decays and data samples.}
{\begin{tabular}{@{}cccccccc@{}} \toprule
Run & System & Trig. & Events & $\int$L & Decays & BR (\%) & $p_{T}$ ($GeV/c$) \\ 
\colrule
Run3 & $p+p$ &$\gamma$ & $4.6\times10^{7}$ & 0.22 $pb^{-1}$ & $\eta\rightarrow\pi^{0}\pi^{+}\pi^{-}$   & $22.7 \pm 0.4$        & $3 - 8$        \\
     &       &                      &                   &                & $\omega\rightarrow\pi^{0}\pi^{+}\pi^{-}$ & $89.1 \pm 0.7$        & $2.75 - 9.25$  \\
     &       &                      &                   &                & $\omega\rightarrow\pi^{0}\gamma$         & $8.9^{+0.27}_{-0.23}$ & $2.5 - 6.5$    \\
     &       &                      &                   &                & $K_{S}^{0}\rightarrow\pi^{0}\pi^{0}$     & $30.7 \pm 0.1$      & $2.5 - 6.5$    \\
\colrule
Run3 & $d+Au$&$\gamma$ & $2.1\times10^{7}$ & 1.5 $nb^{-1}$  & $\eta\rightarrow\pi^{0}\pi^{+}\pi^{-}$   & & $5 - 8$        \\
     &       &                      &                   &                & $\omega\rightarrow\pi^{0}\pi^{+}\pi^{-}$ & & $3.5 - 9$      \\
     &       &                      &                   &                & $\omega\rightarrow\pi^{0}\gamma$         & & $3 - 7$        \\
     &       &                      &                   &                & $K_{S}^{0}\rightarrow\pi^{0}\pi^{0}$     & & $3.5 - 8.5$    \\
\colrule
Run4 & $Au+Au$ & MB    & $7.8\times10^{8}$ & 129 $\mu$$b^{-1}$  &  $\omega\rightarrow\pi^{0}\gamma$ & & $4 - 9$ \\
\colrule
Run5 & $p+p$ & MB                   & $1.5\times10^{9}$ & 0.07 $pb^{-1}$ & $\omega\rightarrow\pi^{0}\pi^{+}\pi^{-}$ & & $2.25 - 13$    \\
     &       &$\gamma$ & $1.0\times10^{9}$ & 2.5  $pb^{-1}$ & $\omega\rightarrow\pi^{0}\gamma$         & & $2.5 - 11$     \\
     &       &                      &                   &                & $K_{S}^{0}\rightarrow\pi^{0}\pi^{0}$     & & $2.25 - 10$    \\
\botrule
\end{tabular}}
\end{table}

\section{Analysis}
As the first step we reconstruct $\pi^{0}\rightarrow\gamma\gamma$ decay by combining EMCal clusters assuming that they are produced by photons originating from the collision vertex. To reduce number of fake $\pi^{0}$ candidates we select clusters with energy $E_{\gamma} > 0.2$~$GeV$ in $p+p$ and $d+Au$ and $E_{\gamma} > 0.35$~$GeV$ in $Au+Au$ collisions.  Shower profile cut is used to reject broader showers having mostly hadronic origin. In $Au+Au$ collisions we use an asymmetry cut $\alpha= \mid E_{1\gamma} - E_{2\gamma} \mid/(E_{1\gamma}+E_{2\gamma})<0.8$ to further reduce background since high-$p_{T}$ combinatorial pairs are peaked at $\alpha=1$ due to the steeply falling spectrum of single photons. Position and width of the reconstructed $\pi^{0}$-peak are parameterized as a function of $p_{T}$. The width of the $\pi^{0}$-peak decreases from 12~$MeV$ at $p_{T}=1$~$GeV$ to 8~$MeV$ at $p_{T}>3$~$GeV/c$. Pairs of photons with invariant mass within two standard deviations of the $\pi^{0}$-peak position and transverse momentum $p_{T}>1$~$GeV/c$ are selected as $\pi^{0}$-candidates. The candidates are assigned the exact mass of the $\pi^{0}$-meson and measured $p_{T}$ of the pair. 

Depending on the decay mode we further combine a $\pi^{0}$-candidates with other particles in the event. In case of $\omega(\eta)\rightarrow\pi^{0}\pi^{+}\pi^{-}$ decay we combine $\pi^{0}$ candidates with any pair of negatively and positively charged tracks assuming them to be $\pi$-mesons. Momenta of the charged tracks are selected to be below 4~$GeV/c$ and 8~$GeV/c$ for $\eta$ and $\omega$ respectively. At higher momentum the charged particle spectra are dominated by decay products of long living particles. More details are described in~\cite{ppg55}. For $\omega\rightarrow\pi^{0}\gamma$ decay reconstructed $\pi^{0}$-candidates are combined with all other photons from the same event with energy $E_{\gamma}>1$~$GeV$. In case of $K_{S}^{0}\rightarrow\pi^{0}\pi^{0}$ decay we combine two $\pi^{0}$-candidates together. However, if the invariant mass of two most energetic photons assigned to two different $\pi^{0}$-candidates is within four standard deviations of $\pi^{0}$-mass then such pair is rejected. This cut suppresses combinatorial background from misidentified $\pi^{0}$'s. To account for $K_{S}^{0}$-meson life time ($c\tau=2.7$ cm) the decay point for every pair is shifted from the collision vertex by the decay length of $K_{S}^{0}$ along the momentum vector of the pair. Finally we recalculate momenta of the $\pi^{0}$-candidates and eventually momentum of the pair. 

\begin{figure}[h]
\begin{center}
\includegraphics[width=0.99\linewidth]{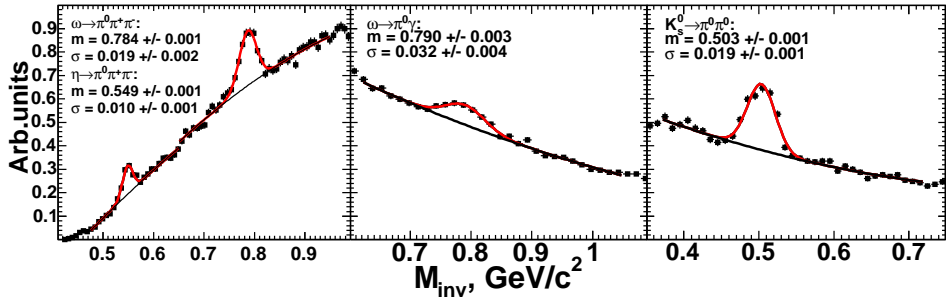}
\end{center}
\caption{\label{fig:minv} Invariant mass spectra for different
%$\pi^{0}\pi^{+}\pi^{-}$, $\pi^{0}\gamma$ and $\pi^{0}\pi^{0}$ 
decays at $5 < p_{T}(GeV/c) < 6$ in $p+p$ collisions.}
\end{figure}

Examples of invariant mass distributions measured in $p+p$ collisions are shown in Figure~\ref{fig:minv}.
Mixed event technique does not reproduce background shape for none of the decays because of particle correlations present in the events. To extract the particle raw yields we use several fitting options. First we fit the peaks with Gaussian function and adjacent background with a parabola as shown in Figure~\ref{fig:minv}. Secondly, the background is approximated in the same region with a parabola, but the points under the peak are rejected from the fit. The parabola interpolated under the peak is subtracted from the histogram. The yield is counted as a sum of the bins in the region initially excluded from the fit. Finally, we narrow the fitting region and use linear function instead of parabolic to describe the background. The three fits are applied to invariant mass spectra collected with different analysis cuts (matching of the charged track to the PC3 hit, tighter cuts for neutral particles) where the shape of the background distribution is different. The raw yields extracted in each case are fully corrected for reconstruction and trigger efficiencies described further. The first of the described fits with its statistical error is taken as the result and a variance of the six measurements is added to the systematic errors as an estimate of yield extraction uncertainty.

For the evaluation of meson reconstruction efficiencies we use Exodus as a particle generator and PISA (PHENIX Integrated Simulation Application) for full Monte-Carlo simulation of the detector and online trigger settings. For the three body decay of $\omega$ and $\eta$-mesons the EXODUS is also used as a particle decay engine weighted to reproduce the phase space density~\cite{phase1,phase2,phase3}. The simulation code is tuned to represent the real configuration of the detector subsystems and to make sure that simulated shape, position and width of the $\pi^{0}$, $\eta$, $\omega$ and $K_{S}^{0}$-peaks are consistent with the ones measured in real data at all $p_{T}$. 
\begin{figure}[h]
\begin{center}
\includegraphics[width=0.49\linewidth]{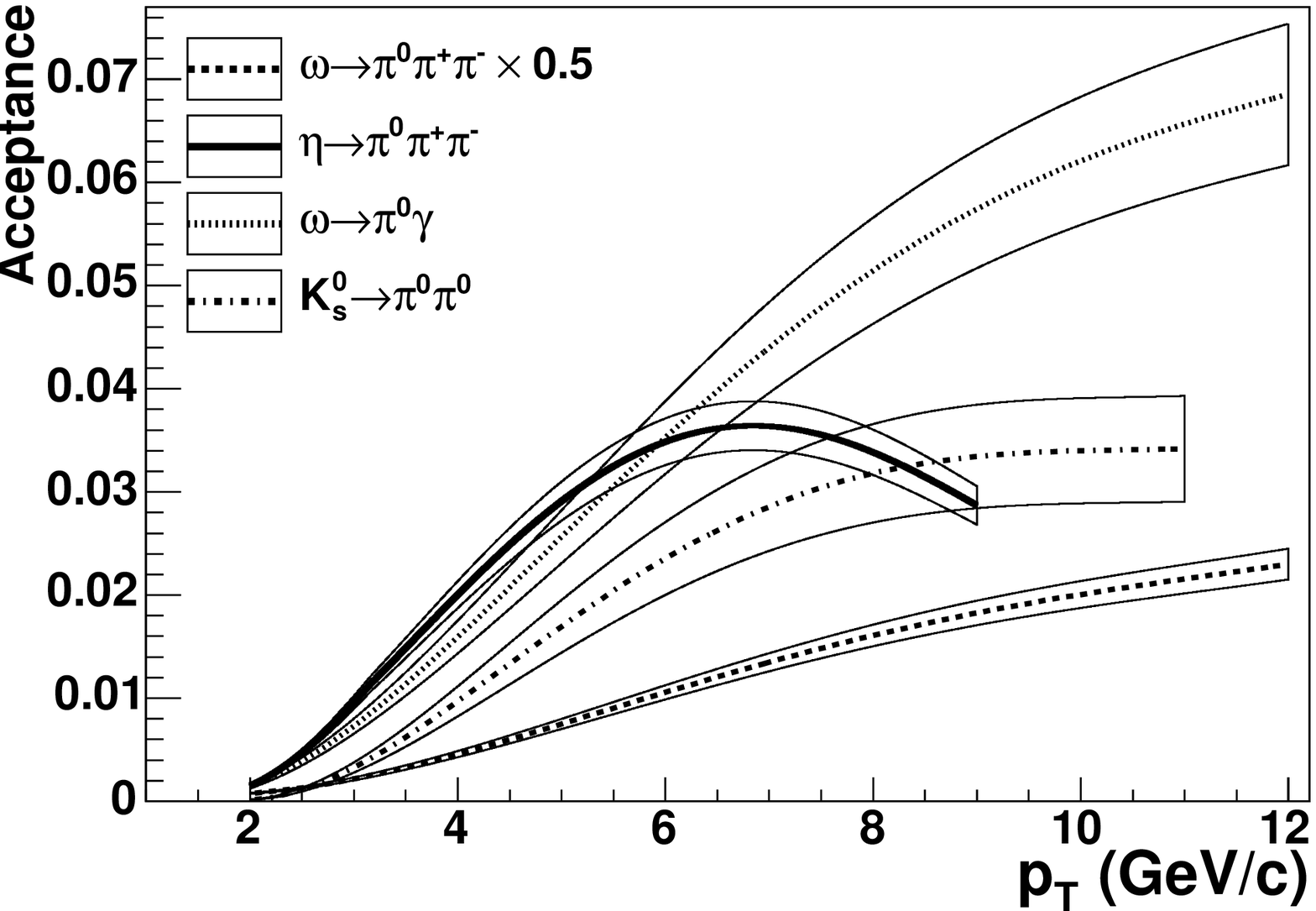}
\includegraphics[width=0.49\linewidth]{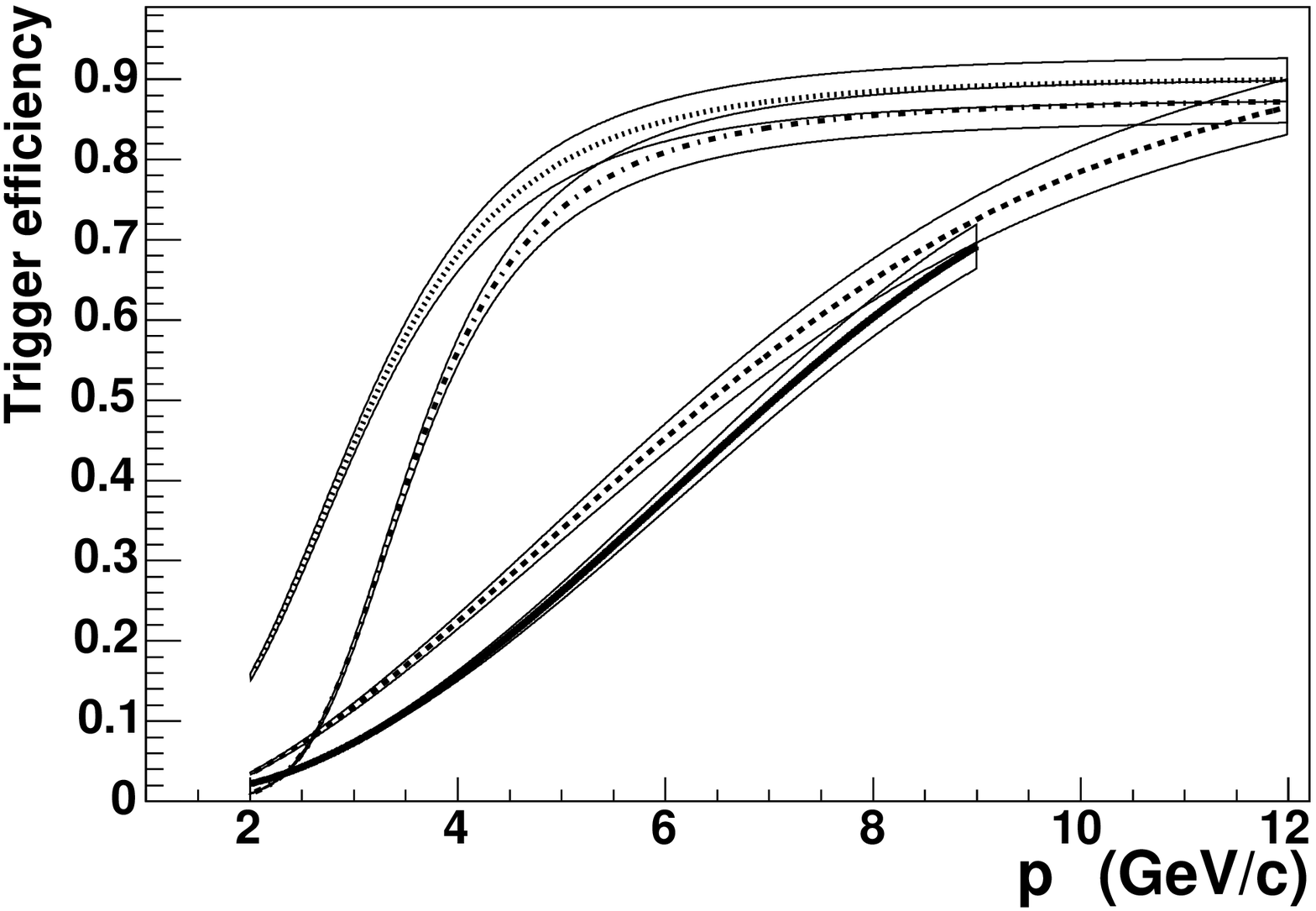}
\end{center}
\caption{\label{fig:sim} Geometrical acceptances and meson trigger
efficiencies for hadron decays of $\omega$, $K_{S}^{0}$ and
$\eta$-mesons. Systematic uncertainties are shown with bands.}
\end{figure}
Left panel of Figure~\ref{fig:sim} shows the acceptance efficiencies which depend on the detector geometry, decay kinematics, performance of detector subsystems and on the cuts used in the analysis. Right panel of Figure~\ref{fig:sim} shows the high-$p_{T}$ $\gamma$-trigger efficiencies for reconstructed decays in $p+p$. They are recalculated from the measured single photon trigger efficiency using kinematics of the particular decays. Both efficiencies must be taken into account to correct the raw yields measured in high-$p_{T}$ $\gamma$-triggered data samples whereas only the curves shown in the left panel are used in MB samples. In $Au+Au$ events one must account for the additional loss in efficiency caused by cluster overlap in the EMCal. This correction is based on simulated single $\omega$-mesons embedded in real events and reconstructed using the analysis procedure. Percentage of the reconstructed mesons is used to correct for this effect.

Systematic errors for different decay modes of $\omega$ and $K_{S}^{0}$-mesons are summarized in Table~\ref{tab:syserr}. Details for $\eta$-meson are described in~\cite{ppg55}. Total systematic error of the measurements is dominated by raw yield extraction, meson trigger efficiency and total cross section uncertainties.
\begin{table}[pt]
\tbl{\label{tab:syserr}Relative systematic errors (\%) for different
decay modes and collision systems. Values with a range indicate
variation of the systematic error over the $p_{T}$ range of the
measurement.}
{\begin{tabular}{@{}lccccccc@{}} \toprule
Source & \multicolumn{2}{c}{$\omega\rightarrow\pi^0\pi^+\pi^-$} 
       & \multicolumn{3}{c}{$\omega\rightarrow\pi^0\gamma$} 
       & \multicolumn{2}{c}{$K_{S}^{0}\rightarrow\pi^{0}\pi^{0}$} \\
& $p+p$  & $d+Au$ & $p+p$ & $d+Au$ & $Au+Au$ & $p+p$ & $d+Au$     \\ 
\colrule
Acceptance             & $5 - 10$  & $9 - 12$   & $10 - 20$ & $8 - 12$ & $14 - 16$ & $10 - 25$ & $10 - 20$ \\
Trigger efficiency      & $3 - 10$  & $5 - 7$    & $2 - 7$   & 5           & -            & $2 - 10$  & 5            \\
Yield extraction       & $5 - 25$  & $10 - 15$  & $5 - 15$  & 10          & $15 - 35$ & $7 - 30$  & 9            \\
MB trigger             & 10           & 8                            & 10          & 8            & 4            & 10           & 8            \\
\colrule
Total                  & $15 - 25$ & $18 - 22$  & $15 - 25$ & $17 - 20$& $20 - 45$ & $20 - 40$ & $15 - 25$ \\
\botrule
\end{tabular}}
\end{table}

\section{Results}
\begin{figure}[h]
\begin{center}
\includegraphics[width=0.49\linewidth]{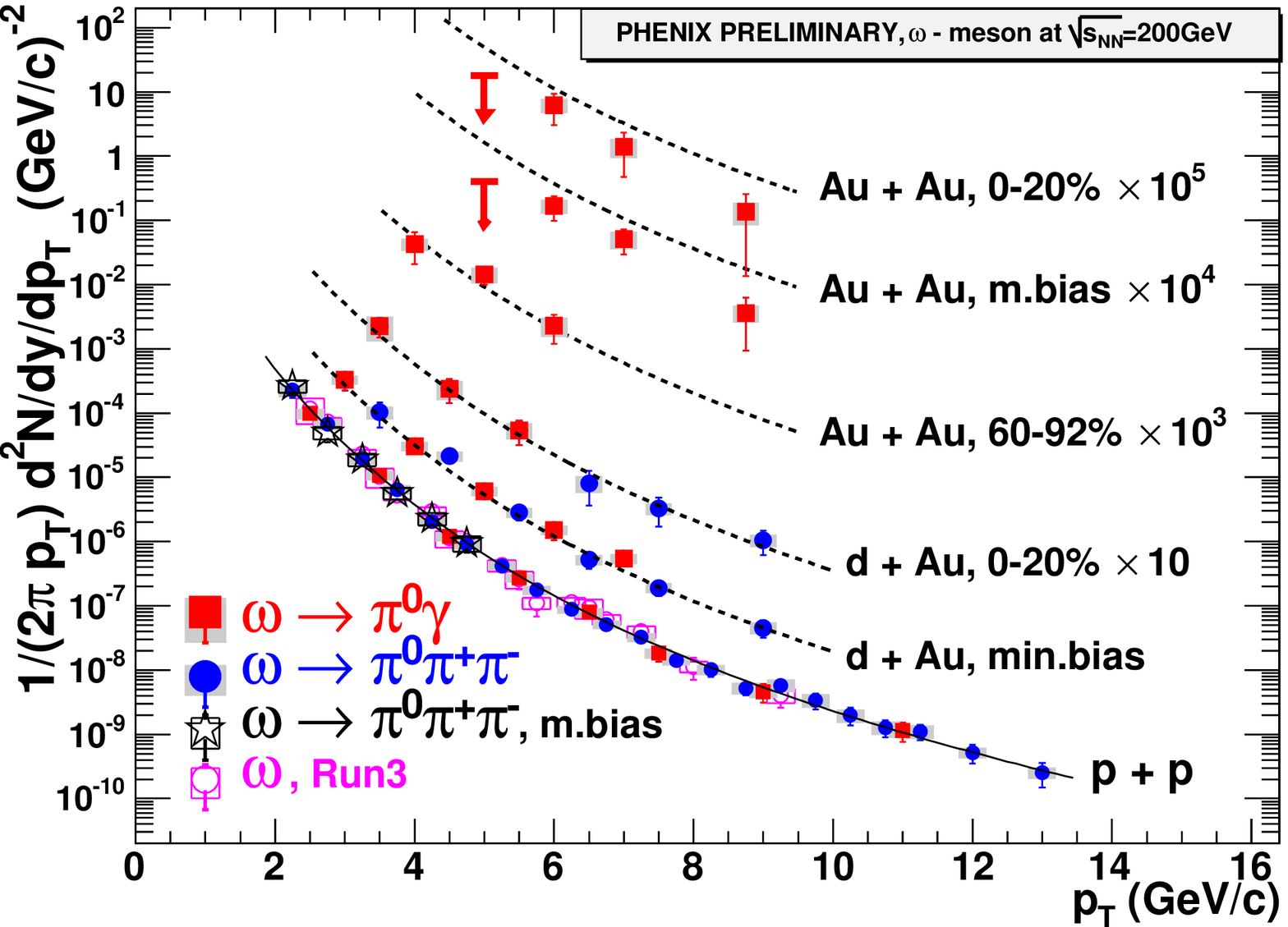}
\includegraphics[width=0.49\linewidth]{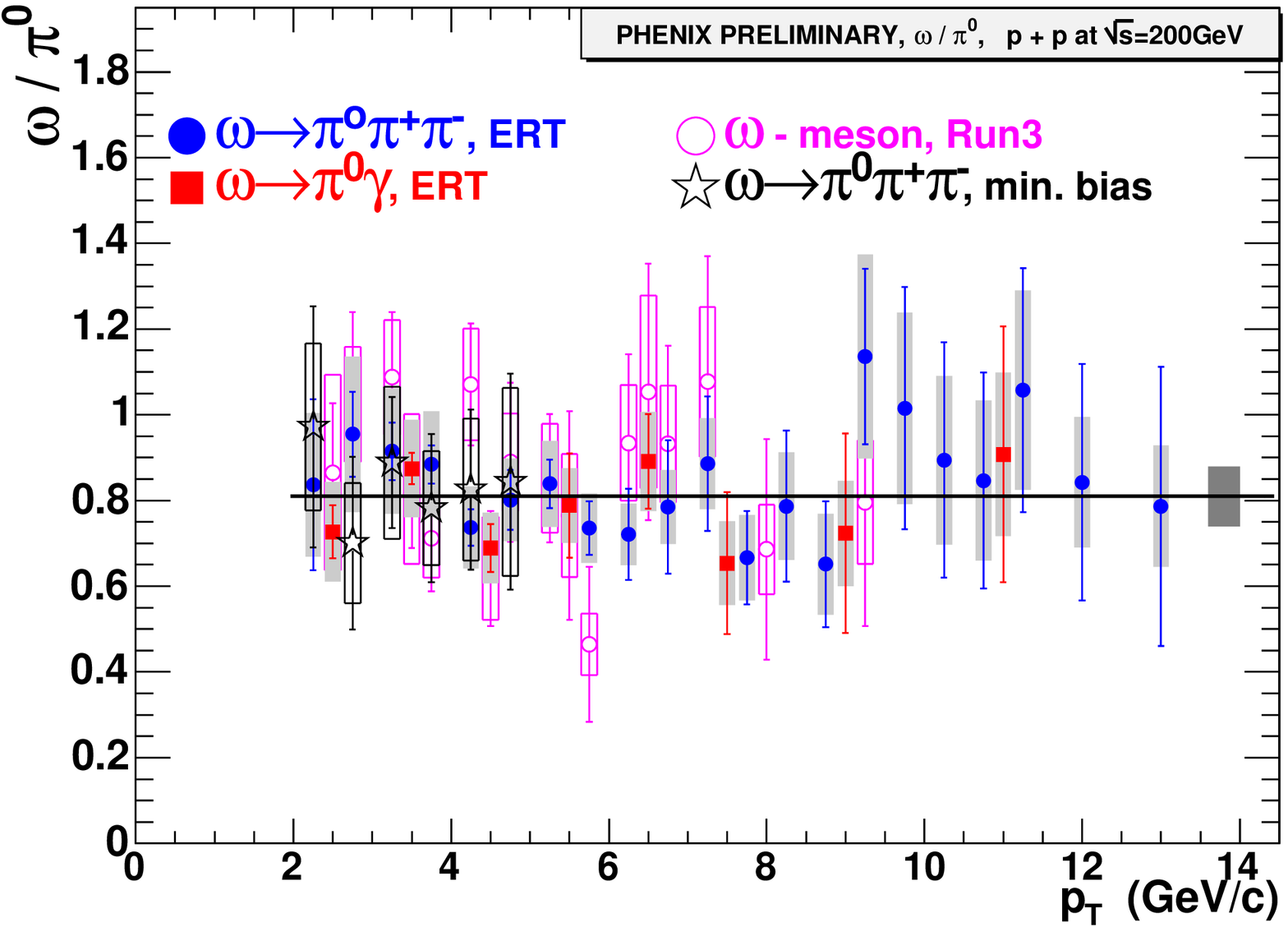}
\end{center}
\caption{\label{fig:omega} Left: Invariant $p_{T}$ spectra for $\omega$-mesons in $p+p$, $d+Au$ and $Au+Au$ collisions at $\sqrt{s_{NN}}=200$~$GeV$. Right: ($\omega$/$\pi^{0}$) ratio as a function of transverse momentum. Statistical errors are shown with error bars and systematic errors are shown with boxes. }
\end{figure}

The invariant $p_{T}$ spectra measured for $\omega$-mesons at $\sqrt{s_{NN}}=200$~$GeV$ are shown in the left panel of Figure~\ref{fig:omega}.
In $p+p$ collisions we observe a very good agreement between Run3~\cite{ppg64,yura} and Run5 results and between the two decay modes measured in the analysis. In Run5 both the high-$p_{T}$ and the MB trigger data give consistent results.  Dashed curves in $d+Au$ and $Au+Au$ points are the fits to $p+p$ spectrum scaled by the corresponding number of binary collisions. As one can see the production of $\omega$-mesons in $d+Au$ and peripheral $Au+Au$ collisions follows the binary scaling, while in the MB and in the most central $Au+Au$ collisions we observe a suppression of $\omega$-meson production with $R_{AA}=0.4\pm0.15$. Ratio of $\omega$-meson and $\pi^{0}$-meson spectra~\cite{neutmes} in $p+p$ collisions is shown in the right panel of Figure~\ref{fig:omega}.  The ratio is flat in the $p_{T}$ range of the measurement. Fit to a constant shown with a solid line gives a value of $0.81\pm0.02\pm0.07$ that agrees with previous PHENIX measurements of $0.85\pm0.05\pm0.09$~\cite{ppg64}. Both measurements give a value of the ratio slightly below PYTHIA~\cite{pythia} prediction of one. Lower energy measurements of R-806~\cite{r806} and E706~\cite{e706} collaborations are consistent with the presented results. 

In Run3 we also measured $\eta\rightarrow\pi^{0}\pi^{+}\pi^{-}$ decay mode. The result is published in~\cite{ppg55}. From the analysis standpoint this decay is very similar to the $\omega$-meson decay and the result is consistent with the $\eta\rightarrow\gamma\gamma$ measured by PHENIX.

\begin{figure}[h]
\begin{center}
\includegraphics[width=0.49\linewidth]{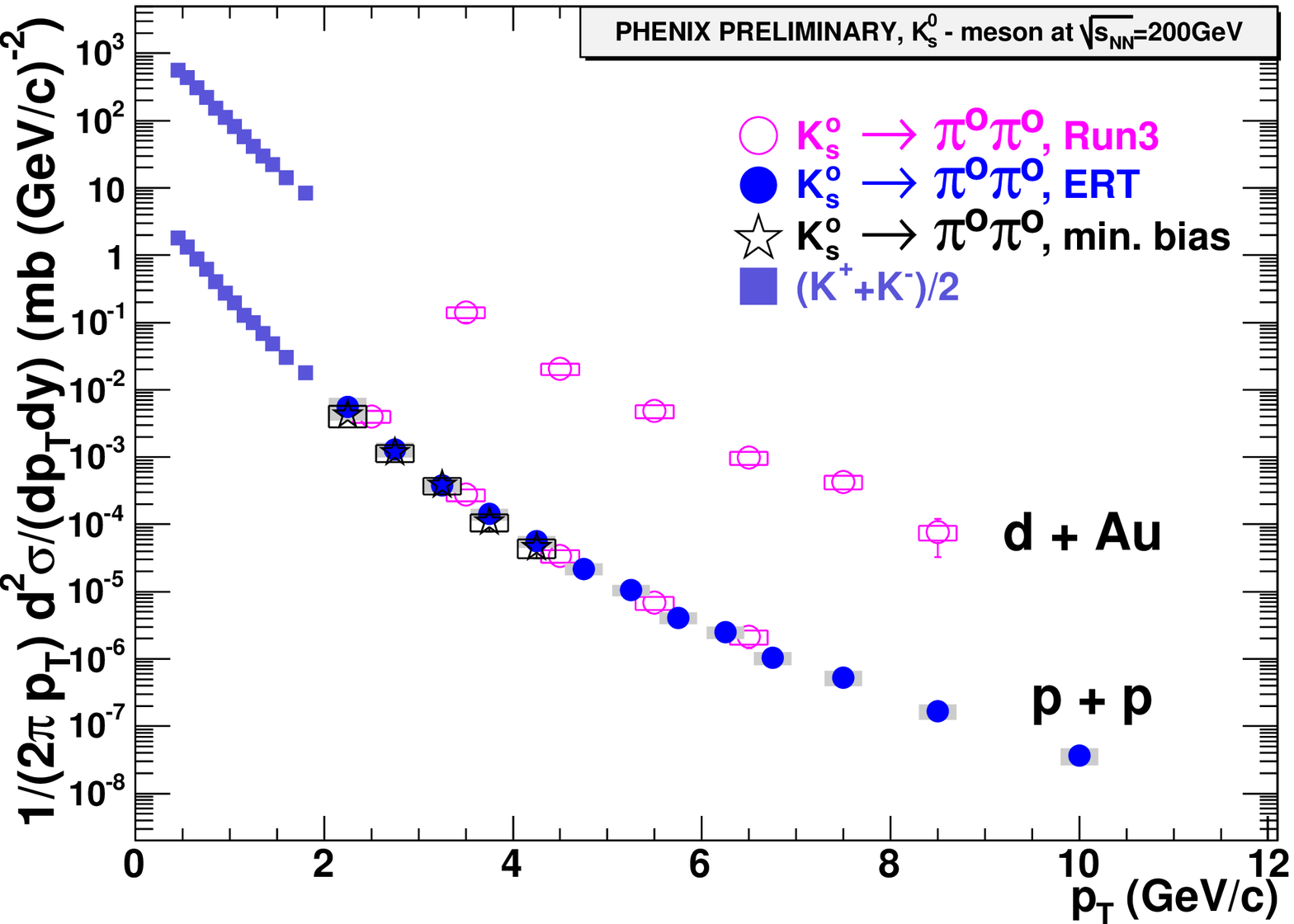}
\includegraphics[width=0.49\linewidth]{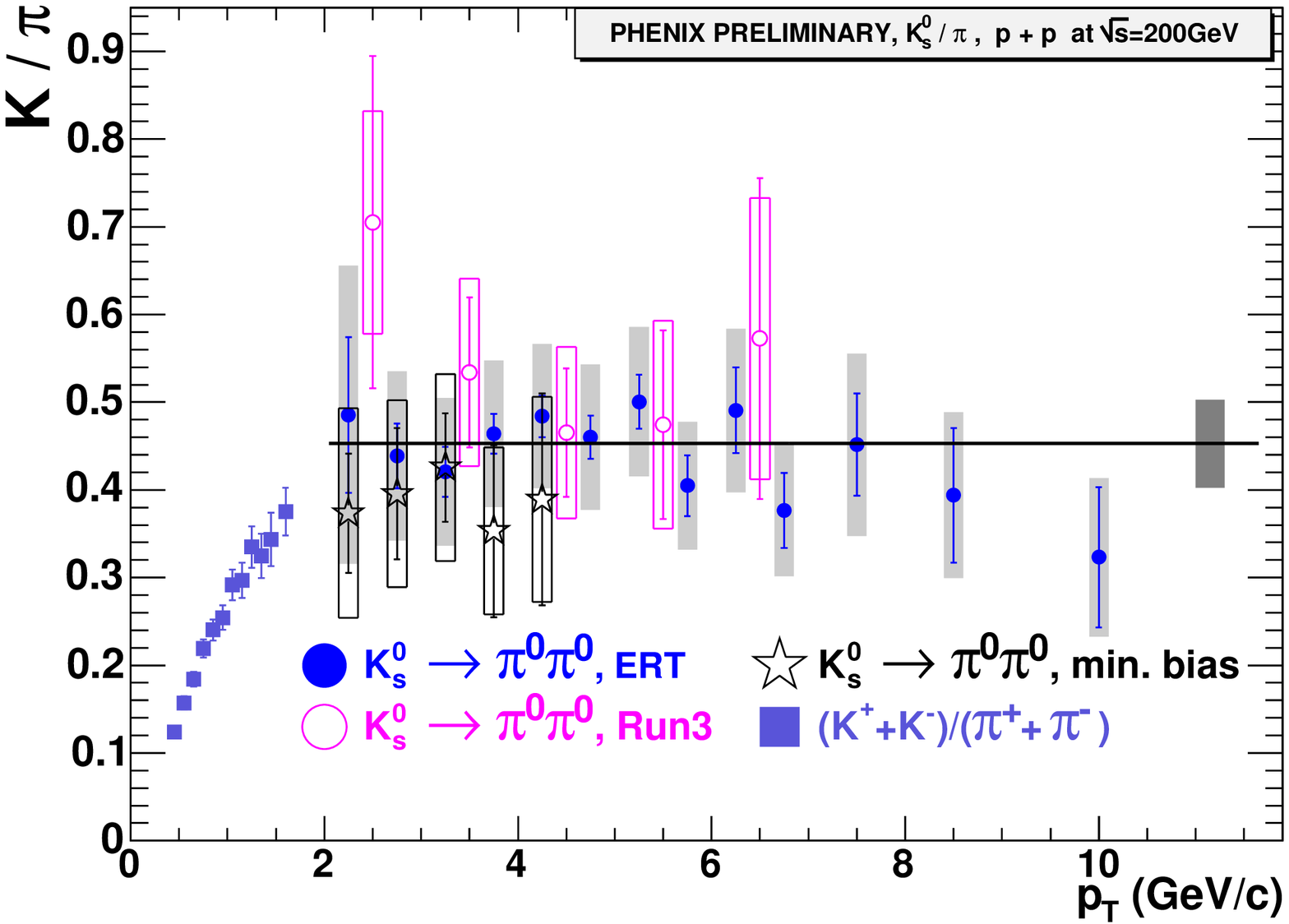}
\end{center}
\caption{\label{fig:ks} Left: Invariant cross section for $K$-mesons in $p+p$ and $d+Au$ collisions at $\sqrt{s_{NN}}=200$~$GeV$. Right: $K$/$\pi^{0}$ ratio as a function of transverse momentum. Statistical errors are shown with error bars and systematic errors are shown with boxes. }
\end{figure}

Invariant cross sections of the $K$-meson production in $p+p$ and $d+Au$ collisions at $\sqrt{s_{NN}}=200$~$GeV$ are shown in the left panel of Figure~\ref{fig:ks}. The $K_{S}^{0}\rightarrow\pi^{0}\pi^{0}$ results are completed with results of charged kaon measurements using PHENIX time-of-flight~\cite{kaon}. The $p_{T}$ range of Run5 measurements extends to 10~$GeV/c$ providing a baseline for heavy ion measurements. Cross section measured in $p+p$ collisions is consistent with $K_{S}^{0}\rightarrow\pi^{+}\pi^{-}$ results from STAR in the overlap region~\cite{starkaon}. Dependence of the $K/\pi$ ratio on transverse momentum is shown in the right panel of Figure~\ref{fig:ks}. Fit to a constant at $p_{T} > 2$~$GeV/c$ gives a value of $0.45\pm0.01\pm0.05$.

\end{document}